\RequirePackage{lineno}
\pdfoutput=1
\documentclass[aps,prl,twocolumn,showpacs,superscriptaddress,groupedaddress,floatfix]{revtex4} 
\usepackage{graphicx} 
\usepackage{dcolumn}  
\usepackage{bm}       
\usepackage{amssymb}  
\usepackage{multirow}
\usepackage{epsfig}
\usepackage{epstopdf}
\usepackage{xspace}
\usepackage{morefloats}
\usepackage{placeins}
\usepackage{amsmath,booktabs,array}
\usepackage[hyperindex,breaklinks]{hyperref}
\usepackage{atlasphysics}
\usepackage{eqparbox}
\usepackage{subfigure}
\def\sww{\ensuremath{W^{\pm}W^{\pm}jj}\xspace}

\newcommand\mjj{\ensuremath{m_{jj}}\xspace}
\newcommand\dyjj{\ensuremath{\vert\Delta y_{jj}\vert}\xspace}

\newcommand{\metpaul}{\mbox{$\rlap{\kern0.15em/}E_T$}}

\newcommand{\WZ}{$WZ/\gamma^*$}

\def\elel{\ensuremath{e^\pm e^\pm}\xspace}
\def\elmu{\ensuremath{e^\pm \mu^\pm}\xspace}
\def\mum{\ensuremath{\mu^\pm \mu^\pm}\xspace}

\usepackage{preprintcover} 

\PreprintIdNumber{CERN-PH-EP-2014-079} 

\PreprintCoverPaperTitle{Evidence for Electroweak Production of \sww\ in $pp$ Collisions at $\sqrt{s}=8$~TeV with the ATLAS Detector}
\PreprintJournalName{Phys. Rev. Lett.}
\PreprintCoverAbstract{This Letter presents the first study of \sww, same-electric-charge diboson production in association with two jets, 
using 20.3~fb$^{-1}$ of proton--proton collision data at $\sqrt{s}=8$~TeV recorded by the ATLAS detector at 
the Large Hadron Collider. 
Events with two reconstructed same-charge leptons (\elel, \elmu, and \mum) and two or more jets are analyzed. Production cross sections are measured in two fiducial regions, 
with different sensitivities to the electroweak and strong production mechanisms. First evidence for \sww production and electroweak-only \sww\ production 
is observed with a significance of $4.5$ and $3.6$ standard deviations respectively. The measured production cross sections are in agreement with Standard Model predictions.
Limits at 95\% confidence level are set on anomalous quartic gauge couplings.}

\begin{document}

\title{\bf Evidence for Electroweak Production of \sww\ in $pp$ Collisions at $\sqrt{s}=8$~TeV with the ATLAS Detector}
\author{The ATLAS Collaboration}
\date{\today}

\begin{abstract}
This Letter presents the first study of \sww, same-electric-charge diboson production in association with two jets, 
using 20.3~fb$^{-1}$ of proton--proton collision data at $\sqrt{s}=8$ TeV recorded by the ATLAS detector at 
the Large Hadron Collider. 
Events with two reconstructed same-charge leptons (\elel, \elmu, and \mum) and two or more jets are analyzed. Production cross sections are measured in two fiducial regions, 
with different sensitivities to the electroweak and strong production mechanisms. First evidence for \sww production and electroweak-only \sww\ production 
is observed with a significance of $4.5$ and $3.6$ standard deviations respectively. The measured production cross sections are in agreement with Standard Model predictions.
Limits at 95\% confidence level are set on anomalous quartic gauge couplings.
\end{abstract}
\pacs{14.70.Fm, 12.60.Cn, 13.85.Fb, 13.38.Be}

\maketitle
 
The scattering of two massive vector bosons (VBS), $VV \rightarrow VV$ with $V=W$ or $Z$, is a key process to probe the nature of electroweak symmetry breaking~\cite{VBF2,VBF3}. In the absence of a Standard Model (SM) Higgs boson, the longitudinally polarized VBS amplitude increases as a function of the center-of-mass energy~$\sqrt{s}$ and 
violates unitarity at energies around 1~TeV~\cite{Unitarity1,Unitarity2,Unitarity3}.
The recent discovery of a 125~GeV SM-like Higgs boson at the Large Hadron Collider (LHC)~\cite{higgs1, higgs2} provides a plausible explaination for the mechanism that unitarizes this process.
However, many physics scenarios predict enhancements in VBS either from additional resonances or if the observed SM-like Higgs boson only partially unitarizes this amplitude~\cite{BSM1,BSM2}. There is no previous evidence for a process involving a $VVVV$ vertex. 

At hadron colliders VBS can be idealized as an interaction of gauge bosons radiated from initial state quarks yielding a final state with two bosons and two jets ($VVjj$) in a purely electroweak process~\cite{VBF1}. VBS diagrams are not separately gauge invariant and must be studied in conjunction with additional Feynman graphs leading to the same $VVjj$ final state~\cite{Gauge_INV}.
Two classes of physical processes give rise to $VVjj$ final states. The first process, which includes VBS contributions, involves exclusively weak interactions at Born level (of order $\alpha_{EW}^{4}$ without considering the boson decay, where $\alpha_{EW}$ is the electroweak force coupling constant) and is referred to as electroweak production. The second process involves both the strong and electroweak interactions at Born level (of order $\alpha_{s}^{2}\alpha_{EW}^{2}$, where $\alpha_{s}$ is the strong force coupling constant) and is referred to as strong production. In the case of same-electric-charge $WW$ production (\sww), the strong production cross section does not dominate the electroweak cross section, making this channel an ideal choice for initial studies on VBS. 

This Letter presents the first evidence for electroweak \sww production, where both $W$ bosons decay leptonically ($W^\pm \rightarrow \ell^\pm \nu$, $\ell=e, \mu$), using $pp$ collision data at $\sqrt{s}=8$~TeV collected by the ATLAS detector at the LHC. This process has a distinct experimental signature of two same-electric-charge leptons and two jets. 

Two fiducial regions are defined.
The first region or ``inclusive region'' is defined to study the combination of electroweak and strong production mechanisms, and in this region both processes are referred to as signal. It is defined at particle level as follows. Exactly 
two prompt charged leptons ($\tau$ leptons and leptons originating from $\tau$ decays are excluded) are required with the same electric charge, 
transverse momentum $\pt>25$~GeV, $|\eta|<2.5$~\cite{coordinate}, invariant mass $m_{\ell\ell}>20$~GeV, and angular separation $\Delta R_{\ell\ell} \equiv \sqrt{(\Delta \phi)^{2}+(\Delta \eta)^{2}}>0.3$.
At least two jets reconstructed with the anti-$k_t$ algorithm~\cite{antikt} with jet size $R=0.4$ and with $\pt>30$~GeV, $|\eta|<4.5$, and separated from the leptons by $\Delta R_{\ell j}>0.3$ are also required. 
The invariant mass of the two jets with the largest $\pt$ ($\mjj$) must be larger than 500~GeV, and the magnitude of the missing transverse momentum~(\met)
 calculated using all neutrinos in the final state must be greater than 40~GeV. 
To reduce the dependence on QED radiation, lepton momenta include contributions from photons within $\Delta R=0.1$ of the lepton direction.
The second region or ``VBS region'' is a subset of the inclusive region that also requires the two jets with largest $\pt$ to be separated in rapidity~\cite{rapidity} by $\dyjj > 2.4$.
This enhances the purity of electroweak \sww by removing most of the strong \sww events, which are considered as a background in this region.

The expected production cross sections for the \mbox{$pp\rightarrow \sww$}\ process
in the two fiducial regions (``fiducial cross sections'')
are calculated using {\sc Powhegbox}~\cite{powhegbox,pb_ssww},
with {\sc CT10} parton distribution functions (PDFs)~\cite{cteq}, 
interfaced with {\sc Pythia8}~\cite{pythia,pythiaAtlasTunes} for 
parton showering, hadronization, and underlying event modeling.
The contribution from non-resonant production of the same leptonic final state is also considered, but is strongly suppressed~\cite{pb_ssww}.
The cross section for the electroweak \sww process 
is predicted to be $1.00 \pm 0.06$~fb in the inclusive region and $0.88 \pm 0.05$~fb in the VBS region. 
The cross section for the strong \sww process is $0.35 \pm 0.05$~fb in the inclusive region 
and $0.098 \pm 0.018$~fb for the VBS region. 
The uncertainty on these predictions include 68\% confidence level PDF uncertainties~\cite{pdfuncert},
parton shower and hadronization modeling uncertainties estimated by comparing {\sc Pythia8} and {\sc Herwig++}/{\sc Jimmy}~\cite{herwig,jimmy},
the independent variation of renormalization and factorization scales by a factor of two,  
the difference between the predictions from {\sc Powhegbox} and {\sc VBFNLO}~\cite{vbfnlo}, and the integration error. 
The parton shower and generator uncertainties are dominant for electroweak production, while scale variations are dominant for strong production.
Interference between electroweak and strong production is studied at leading-order accuracy using {\sc Sherpa}~\cite{sherpa} and is found to 
increase the combined strong and electroweak \sww cross section by $(12\pm 6)$\% in the inclusive region and ($7\pm 4)$\% in the VBS region. 
The total SM signal cross-section prediction in the inclusive region is $1.52\pm 0.11$~fb, while 
the sum of electroweak and interference contributions in the VBS region is $0.95\pm 0.06$~fb.

The ATLAS detector described in Ref.~\cite{atlas} is a multi-purpose particle physics detector. 
It consists of an inner tracking detector (ID) surrounded by a calorimeter and a muon spectrometer (MS). 
Events for this analysis are selected with single-lepton ($e$ or $\mu$) triggers. After applying data quality requirements, the remaining 
data set has a total integrated luminosity of $20.3\pm 0.6$~fb$^{-1}$~\cite{lumi}.

Electron candidates are reconstructed from a combination of a cluster of energy deposits in the electromagnetic calorimeter and a track in the ID.
They are required to have $\pt > 25$~GeV and $|\eta| < 2.47$, excluding the transition region 
between the barrel and endcap calorimeters ($1.37 < |\eta| < 1.52$).
Candidate electrons must satisfy the tight quality definition described in Ref.~\cite{eid} and re-optimized for 2012
data taking. 
Muon candidates are reconstructed by combining tracks in the ID and MS~\cite{muonreco}. The combined track is required to have $\pt > 25$~GeV and $|\eta|<2.4$. 
Leptons are required to originate from the same interaction vertex
and, to reduce non-prompt production, calorimeter and tracker isolation requirements are applied
within a cone of size $\Delta R = 0.3$.

Jets are reconstructed from clusters of energy in the calorimeter using the anti-$k_t$ 
algorithm with jet-size parameter $R = 0.4$ and calibrated using techniques from Ref.~\cite{JES}. 
Only jets with \mbox{$\pt>30$~GeV} and \mbox{$|\eta|<4.5$} are considered. 
Jets containing $b$-hadrons (``$b$-jet'') with \mbox{$|\eta| < 2.5$} are identified by combining information 
on the impact parameter significances of their tracks and explicit secondary vertex reconstruction~\cite{bjet}.
The measurement of \met~\cite{met} is based on the energy collected by the electromagnetic and hadronic calorimeters, and muon tracks reconstructed by the ID and MS. 

Candidate \sww\ events are required to have exactly two leptons (electrons or muons) with the same electric charge
and at least two jets satisfying the above selection criteria. 
Three different final states (``channels'') are considered based on the lepton flavor, namely \elel, \elmu, and \mum.
To reduce the contributions from \WZ+jets and $ZZ+$jets production, 
events are removed if they contain additional leptons reconstructed with looser isolation requirements,
\mbox{$\pt>7$~GeV}~(6~GeV) for electrons (muons) and loose quality definition for electrons~\cite{eid}.
The two leptons must have \mbox{$m_{\ell\ell}>20$~GeV}. 
The dielectron invariant mass must not be within 10~GeV of the $Z$ boson mass to reduce $Z+$jets background from electron charge misidentification.
Events are also required to have \mbox{\met$>40$~GeV}, and in order to reject backgrounds from non-prompt
leptons, mainly $t\bar{t}\rightarrow \ell \nu jjb\bar{b}$, events must not contain a $b$-jet. 
To further reduce $t\bar{t}$ and \WZ+jets backgrounds, events in the inclusive region are required to have \mbox{$m_{jj}>500$~GeV}. 
In addition, in the VBS region \mbox{$\dyjj > 2.4$} is required.

Monte Carlo (MC) simulation is used to estimate the expected signal events.
The \sww\ processes are generated with {\sc Sherpa},
using up to three jets in the matrix-element and parton shower model~\cite{sherpa},
and normalized using the expected cross section in each fiducial region (see above).
Generated events are processed with the full detector simulation~\cite{AtlasSimPaper} based on {\sc GEANT4}~\cite{geant}, and the standard ATLAS reconstruction software.

Several SM processes enter the \sww signal regions as irreducible physics processes or through instrumental effects. 
About 90\% of the expected ``prompt lepton background'' originates from \WZ$\rightarrow \ell^\pm \ell^\mp \ell^\pm \nu$
production that passes signal region selections when one lepton is outside of the experimental acceptance  
or does not satisfy the lepton identification criteria. Up to $20\%$ of the expected \WZ~contribution comes from electroweak production. 
Smaller contributions from $ZZ+$jets and $t\bar{t}+W/Z$ are also considered.
These ``prompt lepton backgrounds'' are estimated using MC simulation. In the VBS region strong \sww is estimated using simulation and normalized to the SM prediction for the fiducial cross section described above. 
Correction factors for lepton and jet efficiencies, additional $pp$ interactions (pile-up), and beam-spot location are applied to the simulation to account for differences with data. 
Furthermore, the simulation is tuned to reproduce the calorimeter response and the muon momentum scale and resolution observed in data. 
Systematic uncertainties on the signal yield and backgrounds estimated from MC simulation are derived from uncertainties on 
the correction factors, energy smearing parameters, the \met\ modeling, and the $b$-tagging efficiency/mistag rate~\cite{bjet}. 
 
{\sc Sherpa} is used to produce \WZ+jets events, taking into account both the strong and electroweak production mechanisms.
This sample is normalized to the next-to-leading-order calculation in QCD from {\sc VBFNLO} in each fiducial region~\cite{vbfnloWZ,Campanario:2013qba}, with an accuracy of 14\% in the inclusive region and 11\% in the VBS region.
The {\sc Sherpa} extrapolation from the inclusive region to the VBS region differs from the {\sc VBFNLO} calculation by $3\%$.
The main sources of uncertainties on the {\sc VBFNLO} normalization are from the PDF, from factorization and renormalization scale dependence, and from the parton shower model.
The small $tZj$ component in this sample is estimated using the {\sc Sherpa} prediction.

The production of $ZZ+$jets is modeled with {\sc Sherpa}, while for $t\bar{t}+W/Z$ processes {\sc Madgraph}~\cite{madgraph}+{\sc Pythia8} is used.
The theoretical uncertainties on the production cross sections of these processes are $\pm19\%$ and $\pm 30\%$ respectively, dominated
by the jet multiplicity modeling and the scale uncertainties. 

Contributions from $W\gamma$ production, including electroweak production of $W\gamma jj$, where the photon converts to an electron--positron pair inside the detector is included in ``conversion background''. It is estimated using 
{\sc Alpgen}\cite{alpgen}+{\sc Herwig/Jimmy} and {\sc Sherpa} (for electroweak $W\gamma jj$) MC samples with a total theory uncertainty of $\pm 17\%$. 

The remaining conversion background originates from processes that produce oppositely charged prompt leptons where one lepton's charge is misidentified, primarily because one electron has undergone hard bremsstrahlung and subsequent photon conversion. This background is estimated from data. The dominant origins of this background are $t\overline{t}\rightarrow \ell \nu \ell \nu b\overline{b}$ and Drell--Yan production. The electron charge misidentification rate is measured using $Z/\gamma^* \rightarrow ee$ events. 
The muon charge misidentification rate is found to be negligible. The background is estimated by applying the electron charge misidentification rate 
to data selected using all signal selection criteria except for the electric charges of the leptons, which are instead required to be opposite sign.
The dominant systematic uncertainties arise from possible method bias (studied in simulation) and the statistical uncertainty in the charge misidentification rate. The total uncertainty is between 15\% and 32\% depending on signal region and channel.

Contributions from SM processes that produce at least one non-prompt lepton from hadron decays in jets 
($W+$jets, $t\bar{t}$, single top or multijet production, 
denoted by ``other non-prompt background'') are estimated from data events that contain one lepton passing all selections and one non-isolated or loose-quality lepton.
These events, which are dominated by the non-prompt background, are scaled by a ``fake rate'' to predict the non-prompt background. 
The fake rate is the efficiency for non-prompt leptons to pass the nominal lepton selections with respect to the looser isolation and quality requirements.
The fake rate for non-prompt leptons is measured in a dijet sample. 
The uncertainty on the non-prompt background estimate is between 39\% and 52\% depending on region and channel, 
dominated by prompt-lepton contamination in the dijet sample and the uncertainty on the extrapolation of
fake rates into the signal region.

Contributions from double parton scattering~\cite{dps} arise mainly in \WZ~and dijet production. However, simulation shows they are negligible after the requirement of $m_{jj}>500$~GeV. 

\begin{table}
\caption{\label{tab:CR}Expected numbers of events (exp.) and measured data counts are shown by channel for each control region described in the text. The uncertainty shown is the systematic uncertainty on the expected yield.}

\begin{tabular}{cc|cccc}
\hline
\hline
\multicolumn{2}{c|}{Control Region} &~Trilepton~ & ~$\leq 1$ jet ~& ~$b$-tagged ~& ~Low $m_{jj}$ ~ \\
\hline
\elel& exp. &  36 $\pm$ 6      & 278 $\pm$ 28 & 40 $\pm$ 6\phantom{0} & 76 $\pm$ 9 \\
&data       & 40    & 288          &  46  &  78  \\
\hline 
\elmu& exp.&   110 $\pm$ 18        & 288 $\pm$ 42  & 75 $\pm$ 13  & 127 $\pm$ 16 \\
&data      &    104      &          328           & 82        &  120   \\
\hline
\mum& exp. &  \phantom{0}60 $\pm$ 10       & \phantom{0}88 $\pm$ 14   & 25 $\pm$ 7\phantom{0} & 40 $\pm$ 6 \\
&data      &      48            & 101          &  36       &  30  \\
\hline
\hline
\end{tabular}
\vspace{-.6cm}
\end{table}

Background predictions are tested in several same-electric-charge dilepton control regions summarized in Table~\ref{tab:CR}.
The MC modeling of prompt backgrounds is tested in a trilepton control region defined by inverting the third-lepton veto and removing the $\dyjj$ and $\mjj$ selections. Conversion and prompt backgrounds are tested in a region with at most one jet (``$\leq 1$ jet'' in Table~\ref{tab:CR}). In this sample the \elel channel is dominated by $Z\rightarrow ee$ events, the \mum channel is dominated by prompt processes, and the \elmu channel has a mixture of prompt, non-prompt, and conversion backgrounds. Backgrounds from non-prompt leptons originating from $t \overline{t}\rightarrow \ell \nu jjb\overline{b}$ are tested in a control region that requires at least one of the jets to be identified as a $b$-jet. Finally, the combined background model is tested by inverting the \mjj selection.

\begin{table*}
\begin{tabular}{l|ccc|ccc}
\hline
\hline
&\multicolumn{3}{c|}{Inclusive Region} &\multicolumn{3}{c}{VBS Region}  \\
& ~~~\elel~~~ & ~~~\elmu~~~ & \mum & \elel & \elmu & \mum\\
\hline
Prompt            & 3.0 $\pm$ 0.7         & 6.1 $\pm$ 1.3         & 2.6 $\pm$ 0.6          & 2.2 $\pm$ 0.5         & 4.2 $\pm$ 1.0        &  1.9 $\pm$ 0.5\\
Conversions       & 3.2 $\pm$ 0.7         & 2.4 $\pm$ 0.8         & --                   & 2.1 $\pm$ 0.5         & 1.9 $\pm$ 0.7        &  -- \\
Other non-prompt  & 0.61 $\pm$ 0.30       & 1.9 $\pm$ 0.8         & 0.41 $\pm$ 0.22        & 0.50 $\pm$ 0.26       & 1.5 $\pm$ 0.6        &  0.34 $\pm$ 0.19\\
\sww Strong      &  0.89 $\pm$ 0.15 &  2.5 $\pm$ 0.4 &  1.42 $\pm$ 0.23  & 0.25 $\pm$ 0.06       & 0.71 $\pm$ 0.14      &  0.38 $\pm$ 0.08 \\
\sww Electroweak &  3.07 $\pm$ 0.30 &  9.0 $\pm$ 0.8 &  4.9 $\pm$ 0.5  &  2.55 $\pm$ 0.25 &  7.3 $\pm$ 0.6  &  4.0 $\pm$ 0.4 \\
\hline
Total background  & 6.8 $\pm$ 1.2         & 10.3 $\pm$ 2.0\phantom{0}        & 3.0 $\pm$ 0.6          & 5.0 $\pm$ 0.9         & 8.3 $\pm$ 1.6        & 2.6 $\pm$ 0.5 \\
Total predicted   & 10.7 $\pm$ 1.4\phantom{0}        & 21.7 $\pm$ 2.6\phantom{0}        & 9.3 $\pm$ 1.0          & 7.6 $\pm$ 1.0         & 15.6 $\pm$ 2.0\phantom{0}       & 6.6 $\pm$ 0.8 \\
\hline
Data              &  12                 &    26               &  12                  & 6                   &     18             &    10       \\
\hline
\hline
\end{tabular}

\caption{Estimated background yields, observed number of data events, and predicted signal yields for the three channels are shown with their systematic uncertainty.
Contributions due to interference are included in the \sww electroweak prediction.}
\label{Tab:summary}
\end{table*}

\begin{figure}[htb]
\begin{center}
\includegraphics[width=0.45\textwidth]{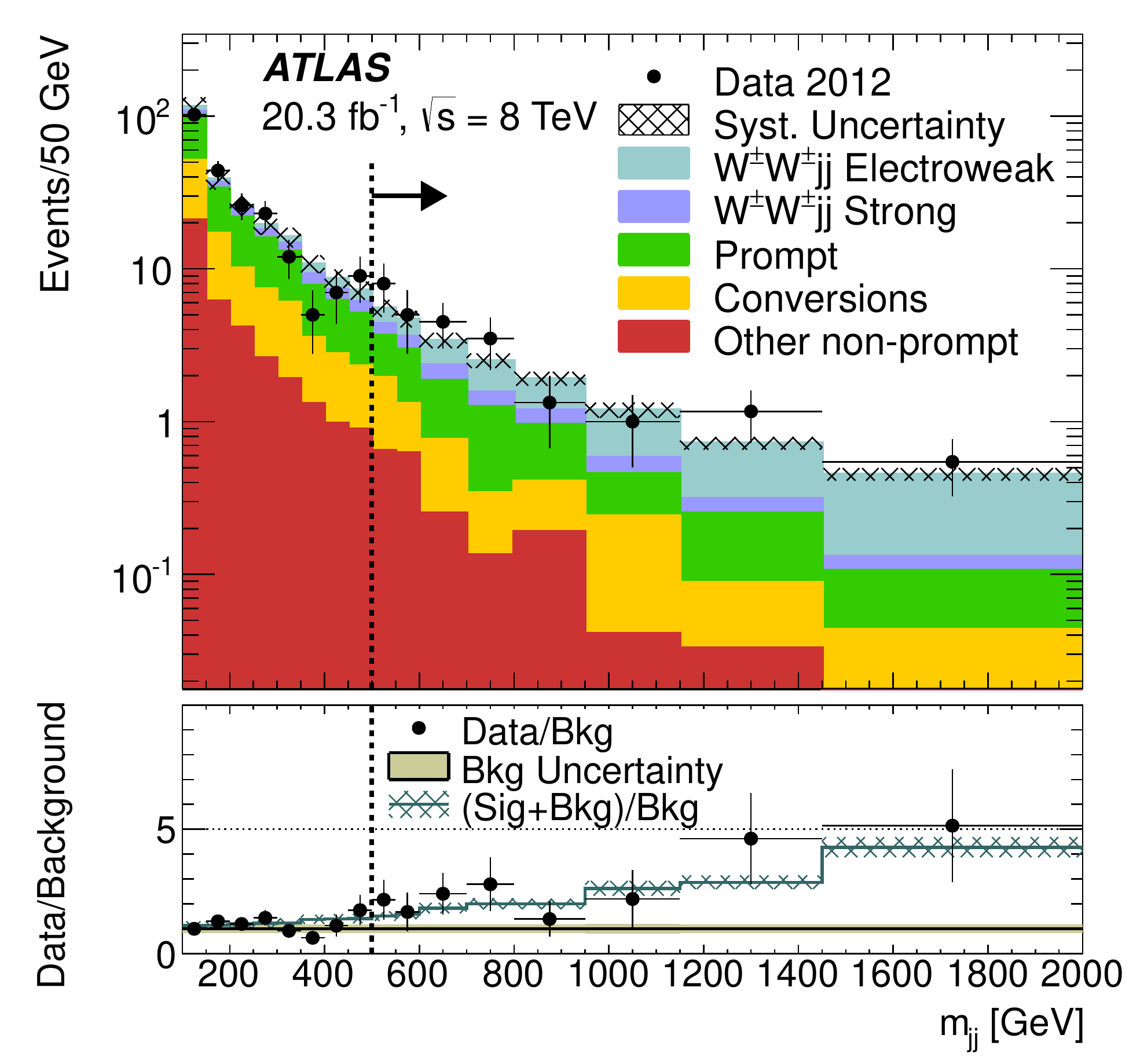}
\caption{The $m_{jj}$ distribution for events passing the inclusive region selections except for the \mjj selection indicated by the dashed line. The black hatched band in the upper plot represents the systematic uncertainty on the total prediction. On the lower plot the shaded band represents the fractional uncertainty of the total background while the solid line and hatched band represents the ratio of the total prediction to background only and its uncertainty. The \sww prediction is normalized to the SM expectation.}
\label{fig:mjj}
\end{center}
\end{figure}

\begin{figure}[htb]
\begin{center}
\includegraphics[width=0.45\textwidth]{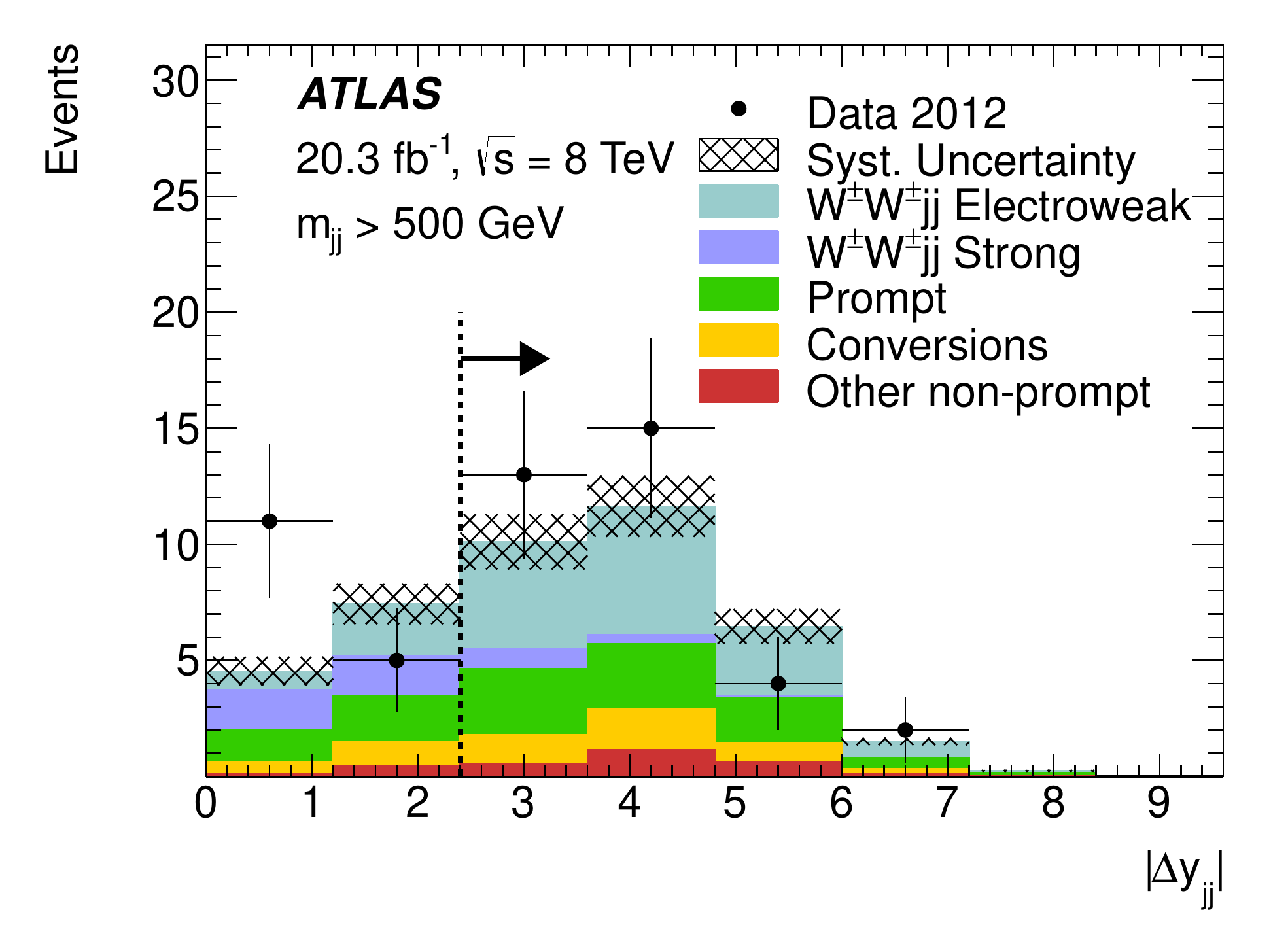}
\end{center}
\vspace{-.4cm}
\caption{The \dyjj distribution for events passing all inclusive region selections. The \dyjj selection is indicated by a dashed line. The \sww prediction is normalized to the SM expectation.}
\label{fig:dyjj}
\end{figure}

The observed number of events is compared in Table~\ref{Tab:summary} to the expected background and
signal yield with systematic uncertainties for the three channels in both the inclusive and VBS signal regions.
In the VBS region strong \sww is considered as background using the SM prediction and its experimental and theoretical uncertainties.
The systematic uncertainty on the background prediction is about 20\%, dominated by the jet reconstruction uncertainties (11--15\%) and
theory uncertainties (4--11\%).
An excess of events over the background expectation is observed in both signal regions and in all three channels;
the combined significance over the background-only hypothesis is 4.5 standard deviations in the inclusive region
and 3.6 standard deviations in the VBS region. 
The expected significance for a SM \sww signal is 3.4 standard deviations in the inclusive region and 2.8 in the VBS region.

Figure~\ref{fig:mjj} shows the expected and observed \mjj distribution after all inclusive region selection criteria are applied, 
except $\mjj > 500$~GeV.
Figure~\ref{fig:dyjj} shows the \dyjj distribution after the inclusive region selections.
All three dilepton channels are summed in both figures. 
The observed excess is consistent with the expected event topology for \sww production.
 
We interpret the excess over background as \sww production, and the fiducial cross sections in the two regions ($\sigma^{\rm fid}$) are measured by combining the three decay channels in a likelihood function. Systematic uncertainties are taken into account with nuisance parameters.

The signal efficiency in each fiducial region is defined as the number of expected signal events after selections
 divided by the number of events passing the respective fiducial region selections at particle level.
The efficiency accounts for the detector reconstruction, migration into and out of the fiducial volume, identification, and trigger efficiency; it is 56\%, 72\%, 77\% for 
the inclusive region and 57\%, 73\%, 83\% for the VBS region in the \elel, \elmu, \mum channels respectively. The efficiency also accounts for the contribution of leptonic $\tau$ decays, which are not included in the fiducial cross-section definition: $10\%$ of signal candidates are expected to originate from leptonic $\tau$ decays. The uncertainty on the signal efficiency is dominated by the jet reconstruction uncertainty of 6\%.
 
The measured fiducial cross section for strong and electroweak \sww production in the inclusive region is $\sigma^{\rm fid}=2.1 \pm 0.5 ({\text{stat}}) \pm 0.3 ({\text{syst}})$~fb. The measured fiducial cross section for electroweak \sww production, including interference with strong production in the VBS region, is $\sigma^{\rm fid}=1.3 \pm 0.4 ({\text{stat}}) \pm 0.2 ({\text{syst}})$~fb.
The measured cross sections are in agreement with the respective SM expectations of $1.52\pm 0.11$~fb and $0.95\pm 0.06$~fb.

\begin{figure}
\includegraphics[width=0.5\textwidth]{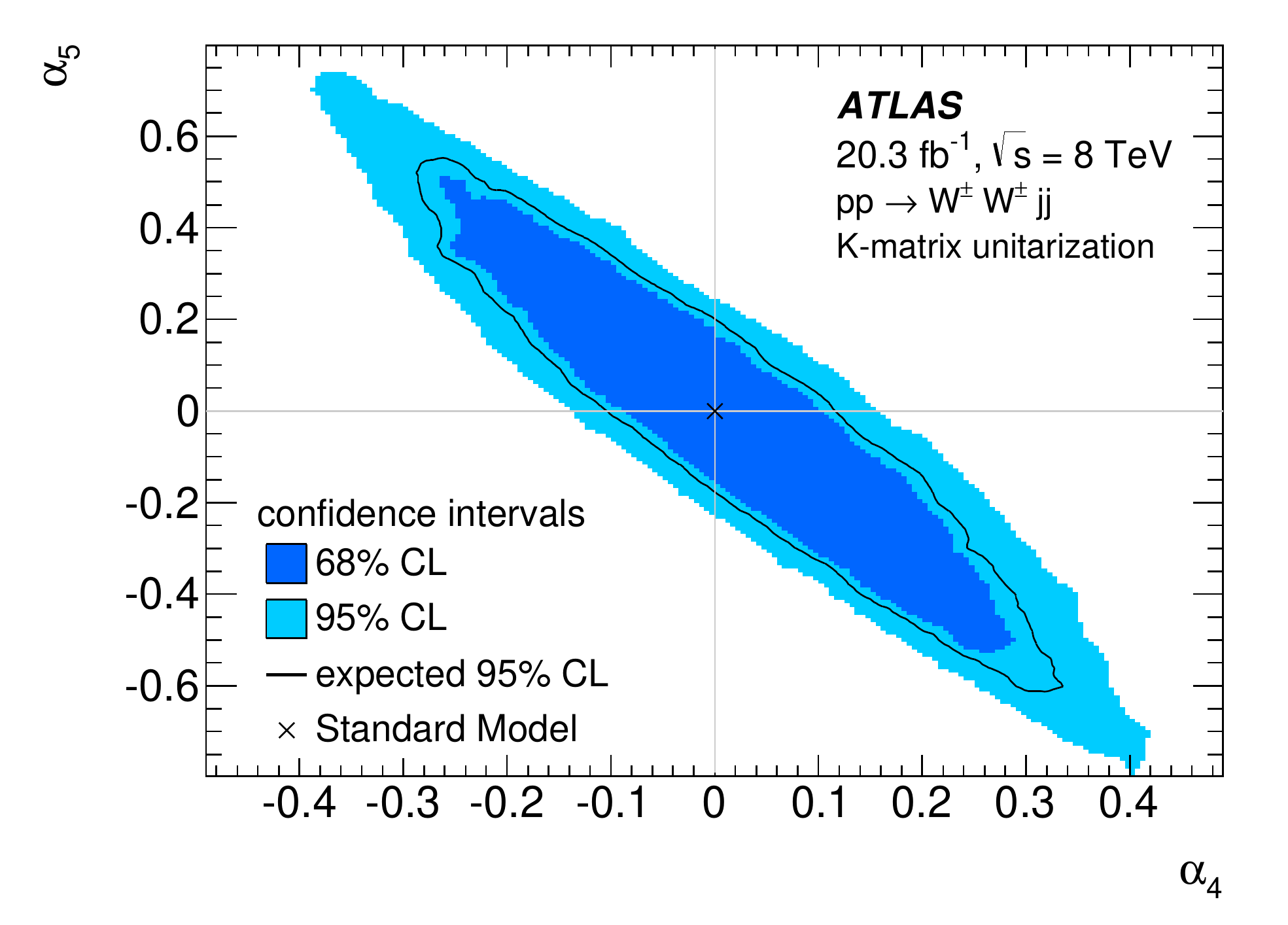}
\caption{\label{fig:aqgc} Limits on ($\alpha_4$,$\alpha_5$). Points outside of the solid light ellipse are excluded by the data at 95\% confidence level (CL). Points outside the inner dark ellipse are excluded at the 68\% confidence level. The expected exclusion is given by the solid line.}
\end{figure}

Additional contributions to \sww production can be expressed in a model-independent way using higher-dimensional operators leading to anomalous quartic gauge boson couplings (aQGCs). 
The measured cross section in the VBS fiducial region is used to set limits on aQGCs affecting vertices with four interacting $W$ bosons.
The {\sc Whizard} event-generator~\cite{whizard} is used to generate \sww\ events with aQGCs 
using a K-matrix unitarization method~\cite{kmatrix}. 
Following existing notations~\cite{kmatrix,aQGC_a4a5}, deviations from the SM (which includes a SM Higgs with $m_H$=126~GeV) are parameterized in terms of two parameters ($\alpha_4$, $\alpha_5$).
The reconstruction efficiency is derived using simulated {\sc Whizard} samples combined with {\sc Pythia8}.
The difference with respect to {\sc Sherpa} for the SM case is taken as additional systematic uncertainty.
The reconstruction efficiency increases with increasing $\alpha_{4,5}$ values, but the effect is small compared to the increase in the fiducial cross sections in the same parameter space.
The expected and observed 95\% confidence intervals derived from the profile likelihood function are shown in Fig.~\ref{fig:aqgc}.
The one-dimensional projection at $\alpha_{5,4}=0$ is respectively $ -0.14 < \alpha_4 < 0.16$ and $-0.23 < \alpha_5 <0.24$ compared to an expected
$ -0.10 < \alpha_4 < 0.12$ and $-0.18 < \alpha_5 <0.20$.

In conclusion, a significant excess of events over background predictions is found using 20.3~fb$^{-1}$ of $pp$ collision data at $\sqrt{s}=8$ TeV recorded by the ATLAS detector at the LHC. This excess is consistent with SM \sww production. Two fiducial cross sections are measured in regions with different sensitivities to the electroweak and strong \sww processes. The measured cross sections are in good agreement with SM predictions. In addition, the first limits on the $\alpha_{4,5}$ aQGC parameters are set.

We thank CERN for the very successful operation of the LHC, as well as the
support staff from our institutions without whom ATLAS could not be
operated efficiently.

We acknowledge the support of ANPCyT, Argentina; YerPhI, Armenia; ARC,
Australia; BMWF and FWF, Austria; ANAS, Azerbaijan; SSTC, Belarus; CNPq and FAPESP,
Brazil; NSERC, NRC and CFI, Canada; CERN; CONICYT, Chile; CAS, MOST and NSFC,
China; COLCIENCIAS, Colombia; MSMT CR, MPO CR and VSC CR, Czech Republic;
DNRF, DNSRC and Lundbeck Foundation, Denmark; EPLANET and ERC, European Union;
IN2P3-CNRS, CEA-DSM/IRFU, France; GNSF, Georgia; BMBF, DFG, HGF, MPG and AvH
Foundation, Germany; GSRT, Greece; ISF, MINERVA, GIF, DIP and Benoziyo Center,
Israel; INFN, Italy; MEXT and JSPS, Japan; CNRST, Morocco; FOM and NWO,
Netherlands; BRF and RCN, Norway; MNiSW, Poland; GRICES and FCT, Portugal; MERYS
(MECTS), Romania; MES of Russia and ROSATOM, Russian Federation; JINR; MSTD,
Serbia; MSSR, Slovakia; ARRS and MVZT, Slovenia; DST/NRF, South Africa;
MICINN, Spain; SRC and Wallenberg Foundation, Sweden; SER, SNSF and Cantons of
Bern and Geneva, Switzerland; NSC, Taiwan; TAEK, Turkey; STFC, the Royal
Society and Leverhulme Trust, United Kingdom; DOE and NSF, United States of
America.

The crucial computing support from all WLCG partners is acknowledged
gratefully, in particular from CERN and the ATLAS Tier-1 facilities at
TRIUMF (Canada), NDGF (Denmark, Norway, Sweden), CC-IN2P3 (France),
KIT/GridKA (Germany), INFN-CNAF (Italy), NL-T1 (Netherlands), PIC (Spain),
ASGC (Taiwan), RAL (UK) and BNL (USA) and in the Tier-2 facilities
worldwide.

\onecolumngrid
\clearpage
\begin{flushleft}
{\Large The ATLAS Collaboration}

\bigskip

G.~Aad$^{\rm 84}$,
B.~Abbott$^{\rm 112}$,
J.~Abdallah$^{\rm 152}$,
S.~Abdel~Khalek$^{\rm 116}$,
O.~Abdinov$^{\rm 11}$,
R.~Aben$^{\rm 106}$,
B.~Abi$^{\rm 113}$,
M.~Abolins$^{\rm 89}$,
O.S.~AbouZeid$^{\rm 159}$,
H.~Abramowicz$^{\rm 154}$
% ...
\bigskip
\\
$^{1}$ Department of Physics, University of Adelaide, Adelaide, Australia\\
% ...
$^{11}$ Institute of Physics, Azerbaijan Academy of Sciences, Baku, Azerbaijan\\
% ...
$^{84}$ CPPM, Aix-Marseille Universit{\'e} and CNRS/IN2P3, Marseille, France\\
% ...
$^{89}$ Department of Physics and Astronomy, Michigan State University, East Lansing MI, United States of America\\
% ...
$^{106}$ Nikhef National Institute for Subatomic Physics and University of Amsterdam, Amsterdam, Netherlands\\
% ...
$^{112}$ Homer L. Dodge Department of Physics and Astronomy, University of Oklahoma, Norman OK, United States of America\\
$^{113}$ Department of Physics, Oklahoma State University, Stillwater OK, United States of America\\
% ...
$^{116}$ LAL, Universit{\'e} Paris-Sud and CNRS/IN2P3, Orsay, France\\
% ...
$^{152}$ Institute of Physics, Academia Sinica, Taipei, Taiwan\\
% ...
$^{154}$ Raymond and Beverly Sackler School of Physics and Astronomy, Tel Aviv University, Tel Aviv, Israel\\
% ...
$^{159}$ Department of Physics, University of Toronto, Toronto ON, Canada\\
% ...
$^{s}$ Also at Novosibirsk State University, Novosibirsk, Russia\\
% ...
\end{flushleft}



\begin{thebibliography}{00}
\bibitem{coordinate}
ATLAS uses a right-handed coordinate system with its origin at the nominal interaction point (IP) 
in the centre of the detector and the $z$-axis along the beam pipe. The $x$-axis points from the IP to the centre of 
the LHC ring, and the $y$-axis points upward. Cylindrical coordinates $(r,\phi)$ are used in the transverse plane, $\phi$ being 
the azimuthal angle around the beam pipe. The pseudorapidity is defined in terms of the polar angle $\theta$ as 
$\eta=-\ln\tan(\theta/2)$. Transverse projections are defined relative to the beam axis.
\end{thebibliography}
\end{document}